\documentclass[preprint]{revtex4}
\usepackage{graphicx}
\usepackage{bm}

\newcommand{\vv}[1]{\mbox{\boldmath{$#1$}}}

\newcommand{\avg}[1]{\left\langle {#1} \right\rangle}
\begin{document}
\title{
Control of electric current by graphene edge structure engineering
}
\author{Masayuki Yamamoto$^1$}
\author{Katsunori Wakabayashi$^{1,2}$}
\affiliation{$^1$ 
International Center for Materials Nanoarchitectonics (MANA),
National Institute for Materials Science (NIMS),
Tukuba 305-0044, Japan
}
\affiliation{$^2$PRESTO, Japan Science and Technology Agency (JST), Kawaguchi
332-0012, Japan}
\date{\today}

\begin{abstract}
In graphene nanoribbon junctions, 
the nearly perfect transmission occurs in some junctions 
while the zero conductance dips due to anti-resonance appear in others.
We have classified the appearance of zero conductance dips 
for all combinations of ribbon and junction edge structures.
These transport properties do not attribute to 
the whole junction structure but the partial corner edge structure, 
which indicates that one can control the electric current 
simply by cutting a part of nanoribbon edge.
The ribbon width is expected to be narrower than 10 nm
in order to observe the zero conductance dips at room temperature.
\end{abstract}

\maketitle

Graphene is one of the most promising materials for future electronics.
The electron mobility of graphene is much larger than that of silicon
and the thermal mobility is twice that of diamond~\cite{emob1,emob2,thmob}.
These properties are, however, 
not enough for fabricating switching devices by graphene.
The electric current must be easily controlled 
by external gate voltage in such devices
while traveling electrons in graphene are hardly stopped 
by potentials due to the Klein tunneling~\cite{Klein,Klein2}.

One possible way for overcoming this problem
is the utilization of edge structures.
In graphene, the electronic band structure of $\pi$ electrons
is strongly affected by edge structures~\cite{peculiar,nakada,prb}. 
Graphene nanoribbons (GNR) with zigzag edges are known
to have partial flat bands near the Fermi energy due to the
edge localized states.
The electronic structures of nanoribbons
with armchair edges crucially depend on the ribbon
width~\cite{peculiar,nakada,prb,son}. 
These properties also result in peculiar transport phenomena
in GNR~\cite{pcc,pcc2,nearlyPCC}.
Moreover, strong energy dependence of conductance has been predicted
when two edge structures are connected at the corner~\cite{junction,ajunc}.

A number of theoretical works have been done on the transport in GNR.
Still, a simple classification of transport properties 
in GNR junctions has been lacked.
Such a classification will be quite useful
for fabricating nanographene devices in the future.
At present, the fabrication of creating clean-edge GNR 
has been experimentally succeeded~\cite{jia,unzip,unzip2}.

In this Letter, we numerically investigate the electronic transport 
in GNR junctions and GNR with partial edge cutting.
In the single-channel energy regime,
the conductance mostly remains unity in armchair edge junction 
while the zero conductance dips appear in zigzag one.
We classify details of the zero conductance dips 
for all combinations of ribbon and junction edge structures.
%

We describe the electronic states of GNR
by the tight-binding model\,,
\begin{equation}
H = -\gamma_0 \sum_{\avg{i,j}} c_i^{\dagger} c_j\,,
\label{hamil}
\end{equation}
where 
$c_i (c_i^{\dagger})$ denotes the creation (annihilation) operator
of an $\pi$-electron on the site $i$, neglecting the spin degree of freedom.
The hopping is restricted in the nearest neighboring atoms.
The conductance is defined by the Landauer formula,
\begin{equation}
g(E) = g_0 {\rm Tr} (\vv{t}^{\dagger} \vv{t})\,,
\end{equation}
with $g_0 \equiv 2e^2/h$.
Here the transmission matrix $\bm{t}(E)$
can be calculated by using the recursive Green function method~\cite{gf}. 

\begin{figure}[ht]
\begin{center}
\includegraphics[scale=0.47]{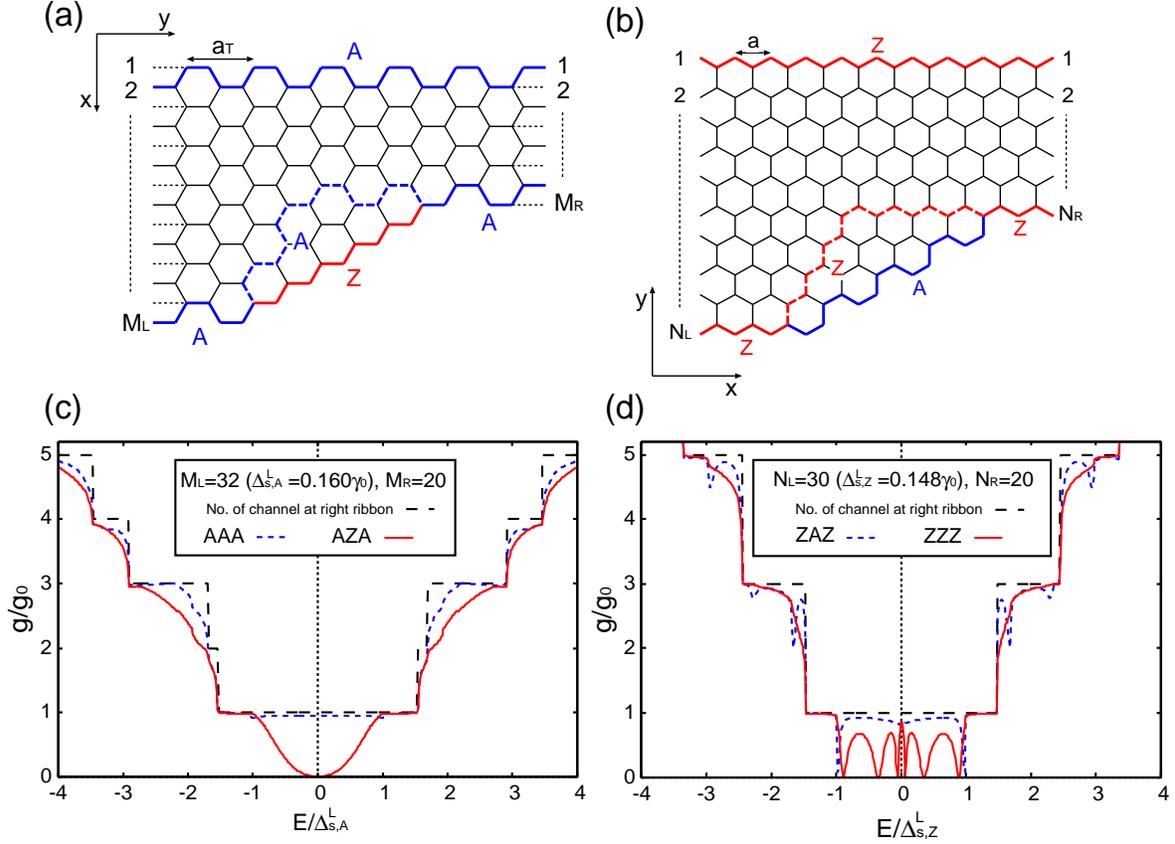}
\caption{
(a) Armchair ribbons with armchair junction (AAA-junction, bule dotted line)
and zigzag junction (AZA-junction, red solid line).
(b) Zigzag ribbons with armchiar junction (ZAZ-junction, blue solid line)
and zigzag junction (ZZZ-junction, red dotted line).
(c) Conductance as a function of energy in armchair ribbon juctions.
The width of left and right ribbons are $M_L=32$ and $M_R=20$.
Conductance mostly remains unity in the AAA-junction
while the zero conductance dip appears in the AZA-junction at $E=0$
for the single-channel energy regime ($|E| < \Delta_{s,A}^L$).
(d) Conductance as a function of energy in zigzag ribbon junctions.
The width of left and right ribbons are $N_L=30$ and $N_R=20$.
Conductance mostly remains unity in the ZAZ-junction
while several zero conductance dips appear in the ZZZ-junction
at non-zero energies
for the single-channel energy regime ($|E| < \Delta_{s,Z}^L$).
}
\label{fig1new}
\end{center}
\end{figure}

Firstly, we consider GNR junctions 
as shown in Fig.\,\ref{fig1new}(a) and (b).
Junctions are classified into
the armchair ribbon with armchair edge junction (AAA-junction),
that with zigzag one (AZA-junction),
the zigzag ribbon with armchair edge junction (ZAZ-junction)
and that with zigzag one (ZZZ-junction).
The width of left (right) armchair ribbon is defined by $M_L (M_R)$
while that of zigzag one is by $N_L (N_R)$.
Here we note that the armchair ribbon is metallic only for
$M_{(L,R)}=3I-1 \,\,(I:{\rm integer})$ 
while the zigzag ribbon is always metallic.

Figure \ref{fig1new}(c) and (d) show the energy dependence of the conductance 
for armchair and zigzag ribbon junctions, respectively.
We set $(M_L, M_R)=(32,20)$ and $(N_L, N_R)=(30,20)$.
The energy is normalized by $\Delta_{s,A}^L\,(\Delta_{s,Z}^L$), 
denoting the end of single-channel energy regime 
of the left wider armchair (zigzag) ribbon. 
These energy scales are related to the ribbon width by
$\Delta_{s,A}^L \simeq \sqrt{3}\pi\gamma_0a/2W$
and
$\Delta_{s,Z}^L \simeq 3\sqrt{3}\pi\gamma_0a/8W$
where 
$W=(M_L+1)a/2$ and $W=\sqrt{3}(N_L+1)a/2$ for armchair and zigzag ribbons,
respectively.
The maximum conductance is limited by the number of channel 
at the right narrower ribbon 
(black dotted line in Fig.\,\ref{fig1new}(c) and (d)).

The behaviour of conductance strongly depends on junction structure
in the single-channel energy regime ($|E| < \Delta_{s,(A,Z)}^L$) while it does 
not in the multi-channel one ($|E| > \Delta_{s,(A,Z)}^L$).
In the single-channel energy regime,
the conductance mostly remains unity in the AAA-junction
while the zero conductance dip appears in the AZA-junction at $E=0$.
This zero conductance dip is due to the anti-resonance 
induced by the coupling between a continuous state at ribbon
and a localized state at zigzag edge junction.
In addition,
the junction region is mainly semiconducting and works as a barrier
for low-energy transport
in the AZA-junction since the ribbon width are narrowed as
$M_L-1, M_L-2, \cdots$.
Hence, by the combination of a resonance and a barrier effect,
the width of zero conductance dip in the AZA-junction
is rather wide and the FWHM (full width at half maximum)
can be roughly estimated as $\Delta_{s,A}^L$.
On the other hand, the junction region is always metallic or semiconducting
in the AAA-junction since the ribbon width are narrowed as
$M_L-3, M_L-6, \cdots$.

In the ZAZ-junction,
the sharp zero conductance dips appear 
in the vicinity of the end of single-channel energy regime
($E \simeq \pm \Delta_{s,Z}^L$).
In zigzag ribbons,
propagating electrons belong to one of two valleys
in the single-channel energy regime
while the second channel will be opened in both valleys
as the energy of incident electrons increases~\cite{pcc2}.
Since the group velocity of a second channel is almost zero 
at the bottom of subband,
the second channel in the other valley works as a bound state 
similar to the zigzag edge state at $E=0$.
The FWHM of dips can be roughly estimated as $\Delta_{s,A}^L/20$
in our numerical simulation performed for several different values of 
the ribbon width $N_L$ and the width difference $\Delta N=N_L-N_R$
within the range $N_L/3 \le \Delta N \le 2N_L/3$.

\begin{figure}[ht]
\begin{center}
\includegraphics[scale=0.5]{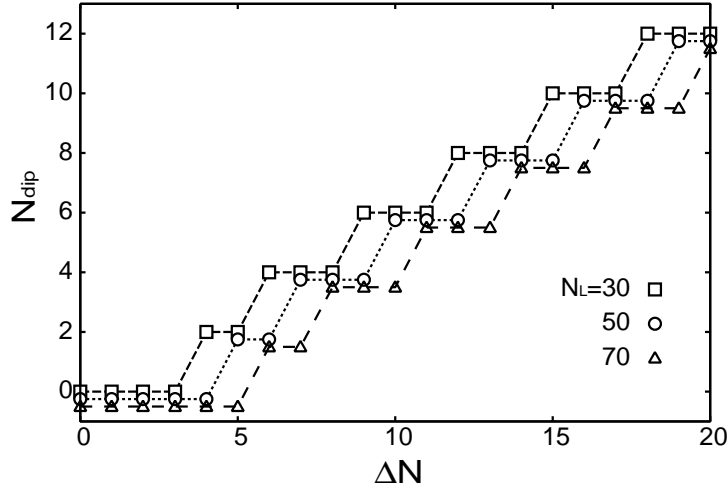}
\caption{
Number of zero conductance dips $N_{dip}$
as a function of the width difference $\Delta N=N_L - N_R$
for $N_L=30, 50$ and $70$ in the ZZZ-junction.
$N_{dip}$ increases by two as $\Delta N$ by three except for small $\Delta N$.
Data for $N_L=50$ and $70$ is slightly shifted downward for legibility.
}
\label{fig2new}
\end{center}
\end{figure}

In the ZZZ-junction,
several zero conductance dips appear at non-zero energies.
%
This is due to the energy level splitting induced by the coupling 
between the edge-localized state on A-sublattice at ribbon
and the edge-localized states on B-sublattice at junction.
Moreover, the coupled states have different nodes
as the width difference $\Delta N$ is getting larger~\cite{junction}.
The number of zero conductance dips $N_{dip}$
is given by $N_{dip}=2(n+1)$ where $n$ denotes the number of nodes.
Figure \ref{fig2new} shows the number of zero conductance dips $N_{dip}$
as a function of the width difference $\Delta N$.
It is shown that, except for small $\Delta N$,
$N_{dip}$ is related to $\Delta N$ as
$N_{dip}=2{\rm Int}\{(\Delta N -N_0(N_L))/3\}$,
where ${\rm Int}\{x\}$ denotes the integer part of $x$
and $N_0(N_L)$ is the integer number depending on $N_L$
(for instance, $N_0(50)=1$ as shown in Fig.\,\ref{fig2new}).
The position and width of each dip depends 
not only on $\Delta N$ but also on $N_L$.
We do not discuss details of them in this Letter,
but the FWHM of the widest dip can be roughly estimated as
$\Delta_{s,Z}^L/5 \sim \Delta_{s,Z}^L/7$
in our simulation.
The appearance of zero conductance dips is classified
for all combinations of ribbon and junction edge structures
in Table \ref{tab1}.

\begin{figure}[ht]
\begin{center}
\includegraphics[scale=0.5]{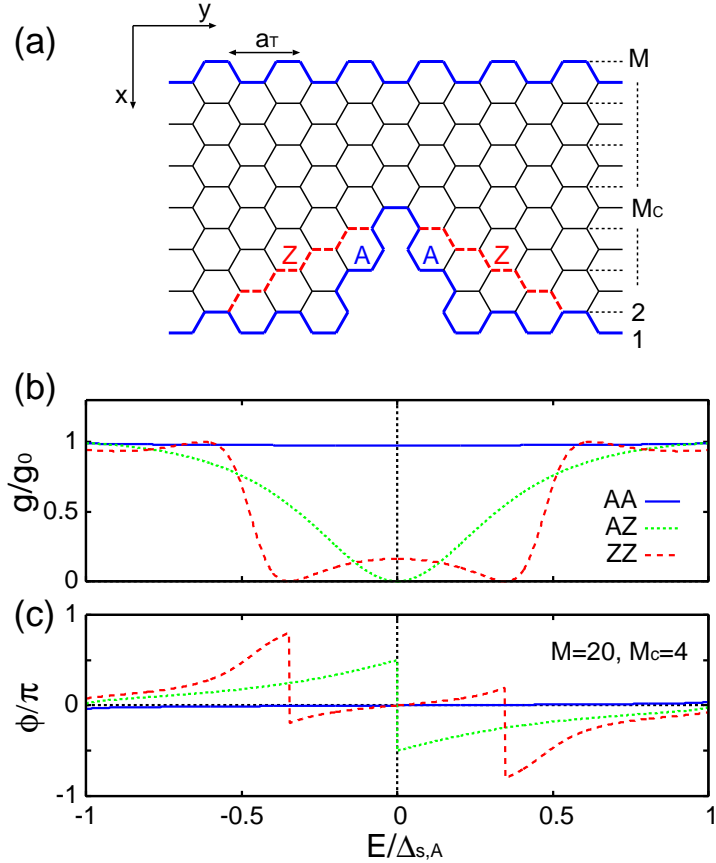}
\caption{
(a) Armchair ribbon with partial edge cutting.
The edge cuttings are classified into armchair-armchair (AA), 
armchair-zigzag (AZ) and zigzag-zigzag (ZZ).
(b) Conductance and (c) phase shift of transmission coefficient
as a function of energy in the AA-, AZ- and ZZ-cuts.
The ribbon width is $M=20$ and the depth of edge-cut is $M_C=4$.
Abrupt $\pi$-phase jump of transmission coefficient
verifies that the appearance of zero-conductance dips is due to anti-resonance.
}
\label{fig3new}
\end{center}
\end{figure}

In order to obtain various zero conductance dips,
it is in fact not necessary to fabricate GNR junctions.
Secondly, we consider an ideal GNR with partial edge cutting
as shown in Fig.\,\ref{fig3new}(a).
The partial edge cutting are classified into armchair-armchair (AA), 
armchair-zigzag (AZ) and zigzag-zigzag (ZZ).
Figure \ref{fig3new}(b) shows the conductance
as a function of energy in the single-channel energy regime
($|E| < \Delta_{s,(A,Z)}$).
Similar to the AAA-junction,
the conductance remains unity for the AA-cut
although it has the sharpest hollow among three cutting patterns.
On the other hand, the zero conductance dips due to anti-resonance 
appear for the AZ-cut and the ZZ-cut.
We note that, in the ZZ-cut,
the number of zero conductance dips increases
as the edge-cut becomes deeper, {\it i.e.}, $M_c$ becomes larger,
as the same as the ZZZ-junction.
This is not case for the AZ-cut where the zero conductance dip
always appears at $E=0$ as the same as the AZA-junction.

In order to clarify that the zero conductance dips are originated from
anti-resonance, we evaluate the phase shift of transmission coefficient
as a function of energy (Fig.\,\ref{fig3new}(c)).
Here the phase shift is given by $\phi=-i\log\left(\vv{t}/|\vv{t}|\right)$
where $\vv{t}$ denotes the transmission coefficient 
in the single-channel energy regime.
It is known that the abrupt $\pi$-phase jump occurs
when the energy of incident electrons are passing through 
the anti-resonance levels~\cite{junction}.
One can clearly see such $\pi$-phase jumps (Fig.\,\ref{fig3new}(c)) 
at the position of zero conductance dips (Fig.\,\ref{fig3new}(b)).

Finally, we discuss the effect of temperature.
At finite temperature, the conductance is given by
$g(E') = g_0 \int {\rm Tr}(\vv{t}^{\dagger}\vv{t}) F_T(E-E')dE$
where $F_T(E)$ denotes the thermal broadening function
$F_T(E) = -\partial f(E) /\partial E$
with the Fermi distribution function $f(E)=1/(\exp(E/k_BT)+1)$.
In order to observe the resonance effect clearly,
the FWHM of zero conductance dips must be wider than
that of $F_T(E)$, which is about $4k_BT$.
The energy width $\Delta_{s,(A,Z)}^L$ is inversely proportional to 
the ribbon width $W$ and one can roughly estimate that 
$W < 12.5$ nm for $\Delta_{s,(A,Z)}^L/4 > k_BT \simeq 300$ K. 
We also note that the phase coherence is necessary in order to 
observe the resonance effect while it can be broken by inelastic 
scatterings due to the electron-phonon coupling at finite temperature.
However, it is known that the effect of electron-phonon coupling is weak 
in carbon nanotube and the inelastic mean free path is estimated as 
$300 \sim 2400$ nm at low bias~\cite{6Suzuura,3Javey,15Park}.
This is due to the peculiar band structure of graphene
so that the same weakness is also expected in GNR.

In conclusion, we have numerically investigated the electronic transport 
in GNR junctions and GNR with partial edge cutting.
In the single-channel energy regime,
the nearly perfect transmission occurs in armchair edge junction while
the perfect reflection occurs at certain energies in zigzag one.
The appearance of zero conductance dips (perfect reflection) is 
due to the anti-resonance between a continuous state at ribbon 
and an localized state at zigzag-edge junction, which can be verified 
by evaluating the phase shift of an transmission coefficient.
We have classified the number, position and width of these zero conductance 
dips for all combinations of ribbon and junction edge structures.
Since these low-energy transport properties do not attribute to the whole 
junction structure but the partial corner edge structure, one can control 
the electric current simply by cutting a part of GNR edge.
Although the ribbon width must be as narrow as 10 nm 
in order to observe this resonance effect at room temperature,
the effect should be detectable at low temperature experiment once the 
technique of fabricating smooth graphene edges is well established.

This work was financially supported by a Grand-in-Aid for Scientific
Research from the MEXT and the JSPS (Nos. 19710082, 19310094 and 20001006). 

\begin{center}
\begin{table}[ht]
\caption{Appearance of zero conductance dips for the combination
of ribbon and junction edge structures.}
\begin{tabular}{ccccc}
\hline
Ribbon & Junction & 
\multicolumn{3}{c}{Zero conductance dips} \\ 
&& Number & Position & Width (FWHM)\\
\hline\hline
A & A & 0 & ----- & ----- \\ \\
\cline{2-5}
& Z & 1 & $E=0$ & $\Delta_{s,A}^L$ \\ \\
\hline
Z & A & 2 & $E \simeq \pm \Delta_{s,Z}^L$　& $\Delta_{s,Z}^L/20$ \\ \\
\cline{2-5}
& Z & 2 Int$\{(\Delta N-N_0(N_L))/3\}$ & *
& $\Delta_{s,Z}^L/5 \sim \Delta_{s,Z}^L/7$ \\ 
&&&& for the widest dip\\
\hline
\multicolumn{5}{l}{A: armchiar, Z:zigzag, *: not discussed in this letter}
\end{tabular}
\label{tab1}
\end{table}
\end{center}


\end{document}